\documentstyle[12pt,epsf]{article}        
\newlength{\dinwidth}                       
\newlength{\dinmargin}                      
\newcommand{\Pomm}{{\tiny \mbox{I$\!$P}}}                
\def\lsim{\mathrel{\rlap{\lower4pt\hbox{\hskip1pt$\sim$}}
    \raise1pt\hbox{$<$}}}                
\def\gsim{\mathrel{\rlap{\lower4pt\hbox{\hskip1pt$\sim$}}
    \raise1pt\hbox{$>$}}}                

\newcommand{\bef}{\begin{figure}}
\newcommand{\eef}{\end{figure}}

\begin{document}
\vspace*{1cm}
\begin{center}  
\begin{Large} 
\begin{bf}
Diffractive structure functions in DIS\\
\end{bf}  
\end{Large}
  \vspace*{5mm}
  \begin{large}
\underline{M.F. McDermott$^a$} and G. Briskin$^{b}$$\footnote{Supported
by MINERVA }$\\ 
  \end{large}
\end{center}
$^a$ Deutsches~Elektronen-Synchrotron~DESY, 
     Notkestrasse~85,~D-22603~Hamburg,~FRG\\
$^b$ School of Physics and Astronomy,
Raymond and Beverly Sackler Faculty of Exact Sciences,\\
\hspace*{1mm} Tel Aviv University,~Israel.\\
\begin{quotation}
\noindent
{\bf Abstract:}
A review of theoretical models of diffractive structure functions
in deep inelastic scattering (DIS) is
presented with a view to highlighting distinctive features, 
that may be distinguished experimentally.
In particular, predictions for the behaviour of the diffractive 
structure functions  
$F_2^D, F_L^D, F_2^{D(\mbox{\small{charm}})}$ are presented. The
measurement of these functions at both small and high values of the
variable $\beta$ and their evolution with $Q^2$ is expected to 
reveal crucial information concerning
the underlying dynamics.

\end{quotation}

\section{Models of hard diffractive structure functions in DIS}

It is natural to start with a definition of what we mean by the
terms `hard' and `diffractive' when applied to scattering of electrons
and protons. High energy scattering processes may be conveniently
classified by the typical scales involved. By hard scattering we mean
that there is a least one short distance, high momentum, scale 
(e.g. high $p_T$-jet, boson virtuality, quark mass) in the problem
that gives one the possibility of using factorization theorems  and
applying perturbative QCD.  In case of diffractive DIS this is the photon
virtuality, $Q^2$, however this hard scale is not necessarily enough
and indeed QCD factorization may not even be applicable to all 
hard diffractive scattering in DIS (see \cite{CHPWW,CFS,BSL} for
discussions and refs).
It has been shown to be applicable to diffractive production of vector
mesons initiated by a longitudinally polarized photon \cite{BFGMS}.
For the time being we will use the definition, 
due to Bjorken, that a diffractive event contains a non-exponentially
suppressed rapidity gap. 
Rapidity is the usual experimental variable related to the
trajectory  of an outgoing particle relative to the interaction point:
given approximately by $\eta \approx  - \ln (\tan (\theta / 2)) $ 
(in a cylindrical system of co-ordinates centered on the
interaction point, with the $z$-axis along the beam pipe and 
polar angle $\theta$).
This rather obscure sounding definition results from the fact that within
perturbative QCD large rapidity gaps (LRG) {\it are} suppressed because a
coloured particle undergoing a violent collision will emit radiation that
would fill up the gap. The suppression factor increases with the  
interval of rapidity but it's absolute magnitude for diffractive
processes in DIS is uncertain. An additional source of rapidity gap
supression comes from an overall damping factor associated
with multiple interactions. The amount of damping is found to be 
much smaller in DIS than that typical for soft processes (e.g. proton
proton collisions see \cite{GLM2}) making LRG events more likely. 


Theoretically, for `diffractive' electron
proton scattering in DIS one must observe the proton in the final state. 
In practice this is very difficult for HERA kinematics since 
the highly energetic scattered proton disappears down the beam pipe in
most events. 
This means that the current measurements also
contain contributions from interactions in which the scattered 
proton dissociates into higher mass states.
This uncertainty is considerably alleviated by the advent of the
Leading Proton Spectrometer
 (LPS) which will provide crucial information about diffraction (for the
first data from the LPS see \cite{zeuslps}). The significance of the 
difference between the experimental working definition of diffraction and the
theoretical one is an interesting but as yet unresolved problem 
(it is certainly possible to produce large gaps in rapidity 
in `non-diffractive' processes, e.g. via secondary trajectory exchanges).


Such LRG events occur naturally in processes known to be
governed by soft processes (e.g proton anti-proton 
scattering at high energies).
This is explained naturally in the context of Regge theory :
at high enough energies one reaches the so-called Regge limit ($s \gg
t$ and $s \gg $ all external masses) 
and {\it all} hadronic total cross sections are expected to be mediated by
Pomeron exchange and to exhibit {\it the same} energy behaviour. This
expectation is born out by the data (see e.g. \cite{DL}), which shows
that a wide variety of high energy total elastic cross sections 
have the same energy dependence which is attributed 
to the trajectory of the soft pomeron. The energy dependence for diffraction
in these processes is discussed in e.g. \cite{GLM3}.


Scattering of virtual photons and protons at small enough $x$ corresponds to
the Regge limit of this subprocess ($\hat s \gg \hat t$, $\hat s \gg
 Q^2, M^2_{\mbox{\tiny{Proton}}}$). It is
natural to ask if the diffractive events observed in the DIS sample
also exhibit the universal behaviour even though we are now considering
off-shell scattering for which, strictly speaking, Regge theory does
not necessarily have to apply. One of the reasons why hard
diffraction at HERA at small $x$ is so interesting is that as $x$
decreases, for fixed large $Q^2$ there should be a transition 
between the hard short distance
physics associated with  moderate values of $x$ and the physics of the
soft pomeron which is widely believed to dominate at very small $x$.
It is a theoretical and experimental challenge to establish whether 
LRG events in DIS in the HERA range are governed by hard or soft 
processes or whether they are actually a mixture of both.
The purpose of this report is to discuss the current theoretical 
models for diffractive structure functions
in an attempt to address this problem, 
and, in particular, to outline the
benchmark characteristics of the various approaches to facilitate the
search for appropriate experimental tests.


In analogy to the total DIS cross section, the diffractive cross
section in DIS can be written,

\begin{equation}
\displaystyle{ 
\frac{ d \sigma^{D} }{ dx_{\Pomm} dt dx dQ^2} = \frac {4 \pi
  \alpha^2_{\mbox{e.m}} }{x Q^4} \left[ 1 - y + \frac{ y^2 }{2 [ 1 + R^{D}
    (x, Q^2, x_{\Pomm},t) ] } \right]  F_2^{D} (x, Q^2, x_{\Pomm},t) 
}
\end{equation}
where $D$ denotes diffraction, $R^D = F_L^D/(F_2^D - F_L^D)$ and
$y=Q^2 / sx$ ; $t=0$ is usually assumed since the cross section
is strongly peaked here.

Ingleman and Schlein \cite{IS} suggested on the basis of expectations from
Regge theory that the diffractive structure functions could be
factorized as follows:

\begin{equation}
{\displaystyle F_2^{D} (x, Q^2, x_{\Pomm},t) = f (x_{\Pomm},t)
    F^{\Pomm}_2 (\beta,Q^2,t)},
\label{eq:fact}
\end{equation}

\noindent where $Q^2$ is the photon
virtuality, $x_{\Pomm}$ is the fraction of the proton's momentum
carried by the diffractive exchange  and  $t$ is the associated
virtuality, $\beta = Q^2 /
(M_x^2 + Q^2 ) = x / x_{\Pomm} $, with $M_x^2$ the mass of the
diffractive system. The last relation for $\beta$ in terms of $x$, the
Bjorken variable, is a good
approximation but only holds for negligible $t$ and proton mass \cite{GS}.
Due to lack of information on the remnant proton both $x_\Pomm$ and 
$t$ can only be estimated indirectly or have to be integrated out.

The 1993 HERA data \cite{h1d2,zeusd5,zeusd4} confirmed the  
presence of events with large rapidity gap between the proton 
direction and the nearest significant activity in the main detector, 
in the total DIS cross section  at the leading twist level
(i.e. this contribution persisted to high values of $Q^2$). These events constitute
approximately 10 \% of the total sample (compared to $\sim 40 \% $ in 
photo-production). As has been known for many years and as Bjorken has recently 
pointed out \cite{bj1} the fact the diffractive cross section 
is present in the total sample as a leading twist effect (i.e. it `scales') 
at large $Q^2$ and small $x$ does not necessarily imply that the mechanism
that creates these events is point-like. For a careful discussion of
the kinematics of hard and soft diffraction in a variety of different
reference frames see \cite{bj2}.

The observed events were also not inconsistent with the Regge
factorization of eq.(\ref{eq:fact}).  Since the cross section had the
same power-like $x_{\Pomm}$ dependence (in $f$) over a the wide range
of ($\beta,Q^2$) that were measured it was tempting to postulate that
a single mechanism or `exchange' was responsible for these events.
The presence of the gap tells us that this object is a colour singlet
and since the centre of mass energy was very high, the exchanged
object became known as the `Pomeron'. From this observation it is
natural to ask, following \cite{IS}, if the partonic content of this
`particle' may be investigated by changing $\beta$ and $Q^2$, with
$\beta$ interpreted as the momentum fraction of the pomeron carried by
the struck parton; $f$ in this picture is interpreted as `the flux of
pomerons in the proton'.


This approach has led to a plethora of theoretical papers in which 
the parton content of the Pomeron  at some small starting scale, $Q_0^2$, 
is treated in various physically motivated ways 
(relying strongly on Regge theory). The DGLAP \cite{DGLAP} equations 
of perturbative QCD (to a given logarithmic accuracy)
are then used to investigate the evolution with $Q^2$ of this parton
content. Formally the use of the DGLAP equations is inapplicable for
the description of diffraction because the presence of the gap makes
it impossible to sum over all possible final hadronic states. Their
use in this context is at the level of a plausible assumption.
In some papers an analogy is drawn with
the proton \cite{GS,GBK1,CKMPT} and a momentum sum rule may be
imposed on the parton content. 
Others models \cite{GS,Kohrs,DL1}
take the view that that the Pomeron may be more 
like the photon and so can have, in addition, a direct coupling to 
quarks within the virtual photon. Although it is no longer clear once
a direct coupling has been introduced whether the concept of a Pomeron
structure function has any meaning. 

\begin{figure}[htb] 
\vspace{2mm}
\begin{center}
\leavevmode
\hbox{%
\epsfxsize=6in
\epsfysize=6in
\epsffile{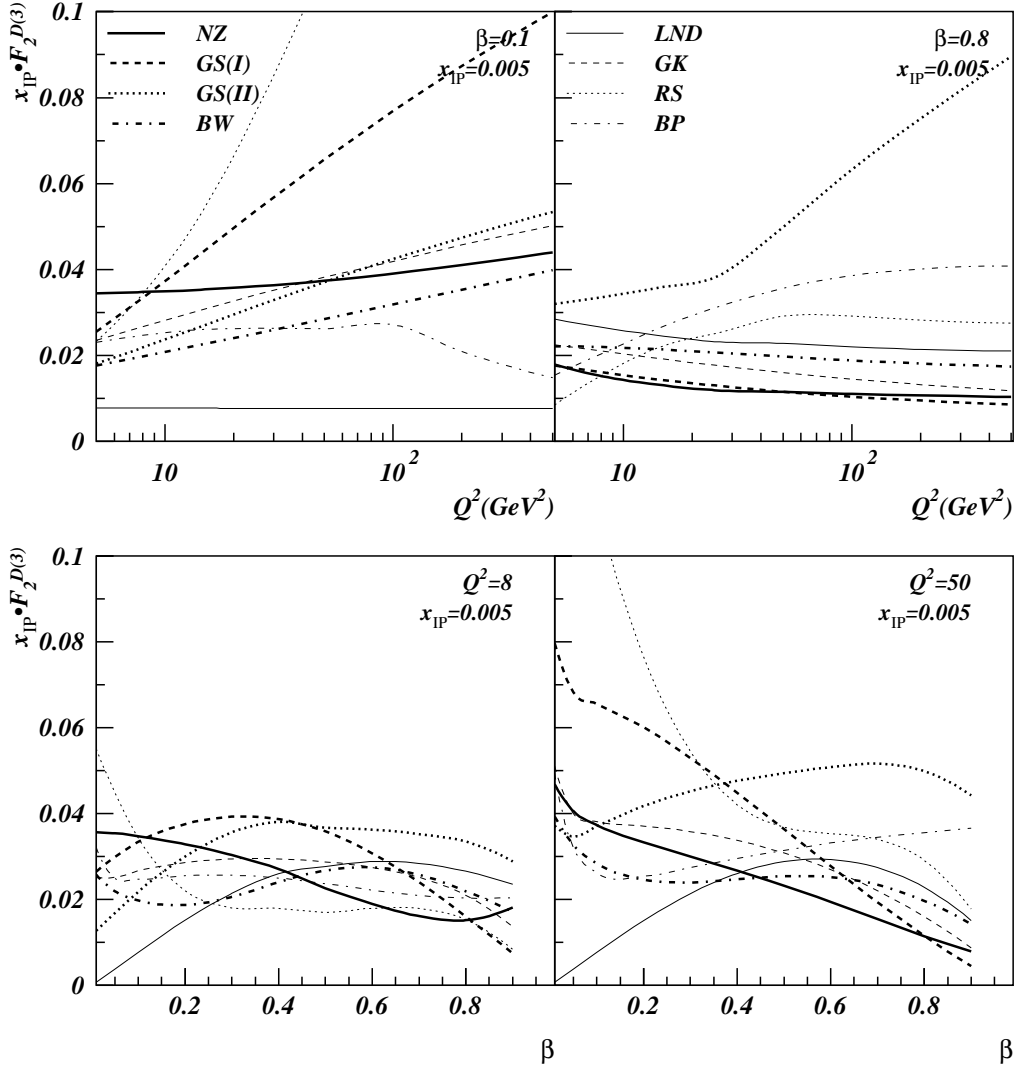}}
\end{center}
\caption[junk]{{\it
Distribution of $x_{\Pomm} F_2^D(\beta, Q^2, x_{\Pomm})$ 
as a function of $\beta$ and $Q^2$, at fixed $x_{\Pomm} = 0.005$,
for various models. For key assignments - see text.}}
\label{fig:F2} 
\end{figure}

Fits \cite{GS,GBK1,JP} to the 1993 data on diffraction reveal 
a partonic structure that is harder (more partons at high $\beta$)
than the proton and that gluons contain a large fraction of pomeron
momentum (up to 90 \%) with a large fraction of these at high $\beta$. 
Clearly in a quantitative sense such statements will depend on the physical
assumptions used to parameterize the input distribution. However
qualitatively these statements are reasonable.
The paper of Gehrmann and Stirling \cite{GS} is 
particularly useful in discussing
Pomeron structure function models in that it discusses and compares
two models: model 1  which has only resolved component and imposes a
momentum sum rule on the parton content and model 2 which also
allows a direct coupling of the Pomeron to quarks.
This leads to
rather different predictions  for the $Q^2$ evolution of these two
models (see curves labelled `GS(I), GS(II)' in fig.(\ref{fig:F2})). 
Model 1 evolves in a way familiar to the
evolution of the proton structure function in QCD, i.e. as $Q^2$
increases there is a migration of partons from high to low $\beta$. 
In model 2, as a result of the direct coupling of the pomeron to quarks
(at `$\beta$ $=1$'), the high $\beta$ distribution is supplemented
and, provided the direct component is large enough, one
expects an increase of parton densities with $Q^2$ over the whole
$\beta$ range, which is also an expectation of the boson-gluon fusion
model of \cite{BH2} (see fig.(\ref{fig:F2})). 

\begin{figure}[htb]  
\vspace{2mm}
\begin{center}
\leavevmode
\hbox{%
\epsfxsize=6in
\epsfysize=6in
\epsffile{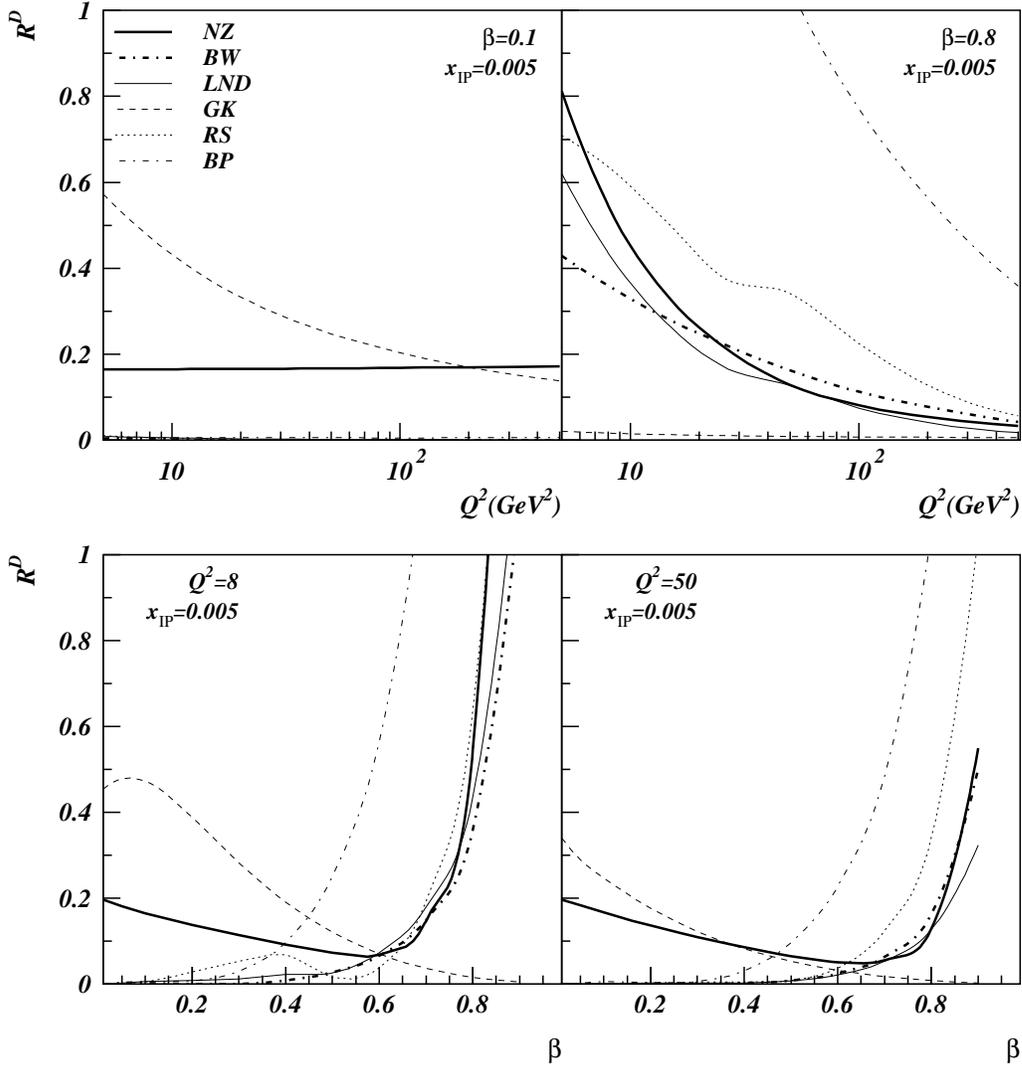}}
\end{center}
 \caption[junk]{{\it $R^D$ as a function of $\beta$ and $Q^2$ for
 $x_{\Pomm}= 0.005$. The pomeron structure function model
 `GK' differs markedly from the two gluon exchange models `NZ,BW,
 LND,RS,BP' at high and low $\beta$. Of these, those based on the dipole
 approach to BFKL, `NZ,BP', produce markedly different $\beta$ and $Q^2$
 behaviour to `RS,BW,LND'.}} 
\label{fig:R} 
\end{figure}

The high gluon content of the pomeron that comes out of the LO QCD
factorizable pomeron models indicate that 
the pomeron structure $R$-factor, 
$R^D(\beta,Q^2,x_{\Pomm}) = 
F_L^D (\beta,Q^2,x_{\Pomm}) / F_T^D (\beta,Q^2,x_{\Pomm})$,  
where $F_T^D = F_2^D  - F_L^D  $, may be
considerably bigger ($R^D \sim 1$) than that for the proton ($R^p
\sim O(\alpha_s)$). Clearly in order to provide a
theoretically consistent prediction for $F_L^D$ a NLO QCD 
calculation is required. Such a calculation has been 
performed by Golec-Biernat and Kwie\'{c}inski
\cite{GBK1} who consider a model with resolved partons in the pomeron
subject to a momentum sum rule. For high $\beta$, $R$ is small in such
models but it can reach 0.5 for $\beta <
0.1$. It has a much softer dependence on $\beta$ than 
$F_2^D$ (see `GK'  in figs.(\ref{fig:F2},\ref{fig:R})).

This picture of the pomeron structure function has been discussed 
in detail elsewhere and will not be repeated in further detail here.
For a lucid account of this picture and of the 1993 data see \cite{JP}.  
The latest results from  H1 \cite{h1d3,h1d4} on the 1994 data (which has a
factor of 10 increase in statistics and covers a broader kinematical
range) suggest that single particle factorization no longer holds over the 
full kinematical range and that particularly for small $\beta$ it
breaks down, i.e. $f$ in eq.(\ref{eq:fact}) become $\beta$ (but not
$Q^2$) dependent. A possible explanation of this is that sub-leading
Regge exchanges play an important r\^{o}le \cite{h1d3,h1d4,GBK2}.

The paper by Ellis and Ross \cite{ER} calls into question the validity of
these parton model approaches using kinematical arguments concerning
the virtuality of the struck parton. They stress the importance 
of measuring diffractive events at high $\beta$
and predict a slow power-like increase with $Q^2$ in this region in
contrast to the logarithmic decrease that may expect from a 
naive QCD evolution.

This and others models are, broadly speaking, similar in spirit 
to the old aligned jet model (AJM), which is a kind of parton 
model approximation to the wavefunction of the photon 
(see \cite{bj1,bj2} and refs.), and
it's QCD improved formulation (see \cite{afs} and refs.). 
Consider virtual-photon proton scattering at high
energies (small $x$) in the proton's rest frame. In this frame the
virtual photon, whose energy, $q_0$, is the largest scale, fluctuates
into a $q {\bar q}$ at a large distances, $l_c = 1/2M_p x = q_0/Q^2$, 
from the proton. 
As  Ioffe \cite{Ioffe} observed many years ago these large
distances are important in determining the DIS structure functions.
For the HERA energy range this `coherence length' can be
as large as 1000 Fm. 
In other words, at enough high energies we may  consider DIS  
as the interaction of the quark anti-quark pair with the proton. The
transverse size of the pair on arrival at the proton is $b_T^2 \approx 1/k_T^2$.

In the configuration in which one of the quarks carries most
of the momentum of the photon a large transverse distance develops
between the fast and the slow quark by the time it arrives at the
proton. This large system, in which the pair is initially `aligned' along the
direction of the original photon, 
essentially interacts with the proton like a hadron. 
This aligned configuration gives a leading twist 
contribution to $F_2$ and $F_2^D$, the latter being interpreted as
the fraction of events where the produced pair is in a colour
singlet state.  Since the slow quark is almost on shell, the AJM is
similar to the parton model and there is no leading twist contribution
to $F_L$ from this configuration.

In the configuration in which the momentum is shared more
equally the quarks can stay closer together in transverse space and may
interact with the proton perturbatively. These configurations 
contribute at leading twist to $F_2(x,Q^2)$ and $F_L(x,Q^2)$. 
In the former the integration over the
momentum fraction leads to the logarithm in $Q^2$ (coming directly
from the box diagram). 
For such small configurations colour transparency phenomena are
expected: the emission of initial
and final state radiation is suppressed \cite{afs}. 

A semi-classical calculation \cite{BH1,BHM} 
in which the proton is treated like a classical background field,
leads to results very similar to those of the AJM. 
Working in the proton's rest frame, one
considers the interaction of different kinematical 
configurations of the highly energetic partons 
in the virtual photon with the soft 
colour field of the proton. These interactions induce non-abelian 
eikonal factors in the wavefunctions of the partons which can lead to
diffractive final states. In \cite{BHM} the addition of gluon 
to the final state is considered. Leading
twist diffractive processes appear when at least one of the three
partons has a small transverse momentum and carries a small fraction
of the longitudinal momentum of the proton. The other two partons may
have large transverse momentum, this means they stay close together as
they move through the proton, acting effectively as one parton. This high
$k_T$ jet configuration, produces the only leading twist contribution
to $F_L^D$ at this order (which is constant) and $\ln {Q^2}$ enhancement of
$F_2^D$. This signals that $F_2^D$ also has leading twist contribution
from the configuration in which all the transverse momenta are small. 
Several qualitative phenomenological predictions come out of this
picture. One expects the ratio $F_2^D/F_2$ to decrease like $\ln {Q^2}$
and there to be fewer high-$p_T$ 
jets in $F_2^D$ than in $F_2$ (they appear only
at order $\alpha_s$ in the former). Leading
twist diffraction appears at order $\alpha_s$ in $F_L^D$ which will
be dominated by jets.

Buchm\"{u}ller and Hebecker \cite{BH2} present a model of diffraction in
DIS based on the dominant process being boson gluon 
fusion, with the colour singlet state being formed by soft 
colour interactions (SCI). The main point is that
diffractive and non-diffractive events differ only by SCI, the
kinematics are expected to be similar since one gluon carries
most of the momentum of the exchanged system.
This idea has also been developed in \cite{ERI1,ERI2} which provides a
Monte Carlo simulation of SCI.

The simplest QCD model for diffractive exchange is a pair of
$t$-channel gluons in a colour singlet state. Such an exchange is a
common feature of many models \cite{BW1,MD,LMRT,LN,GLM,PB1,NZ1} and leads
to a diffractive structure functions which are proportional to the
gluon density squared. The dynamical content of these models differ in
the treatment of QCD corrections and choice of gluon density and will
be discussed in more detail below.

It may be possible to distinguish these models from those 
in which soft colour interactions play a r\^{o}le \cite{BH1,BH2,BHM}
by comparing $F_2^D (x,Q^2,x_{\Pomm})$ with $F_2(x_{\Pomm},Q^2)$ for
fixed $Q^2$ and intermediate $\beta$. For the latter
the following scaling relation is predicted:

\begin{equation}
 F_2^D(x,Q^2,x_\Pomm) \simeq \frac{C}{x_{\Pomm}} F_2(x_{\Pomm},Q^2)
 \label{eq:scaling}
\end{equation}
 
\noindent where $C$ is a constant.

In \cite{BW1} where diffraction is governed by two gluon exchange 
one expects this behaviour to be  multiplied by a
factor $x_\Pomm^{-\lambda}$, where $\lambda \geq 0.08$ and will
depend on $k_T^2$  (see below). 
In the dipole approach  to BFKL \cite{PB1,NZ1}, in which the dipoles
couple via two gluon exchange a similar result is expected but with  
a larger power 
$x_\Pomm^{-\Delta}$, 
$\Delta  \equiv \alpha_\Pomm -1 = 12 \alpha_s \ln(2) / \pi $ possibly softened
by inverse powers of logarithms in $1/x_\Pomm$ \cite{PB1,PB2}.
Of course, in this case, the individual energy dependences 
of $F_2$ and $F_2^D$ is expected to be a lot harder.

In the perturbative QCD approach advocated by  Bartels and 
W\"{u}sthoff \cite{BW1}
the coupling of the pomeron to the hadronic final state can be derived
without any additional parameters except the strong coupling.
The following ansatz is used for the unintegrated gluon density:

\begin{equation}
\psi(x,k^2_T, Q_0^2) \sim \frac{1}{k_T^2 + Q_0^2} x^{1-\alpha_{\Pomm}
  (Q^2)},
\label{eq:bw} 
\end{equation}

\noindent with the effective scale-dependent pomeron intercept (which
explicitly, albeit mildly, breaks the factorization of
eq.(\ref{eq:fact}) since it depends of $Q^2$)
$\alpha_{\Pomm} (Q^2) = 1.08 + 0.1 \ln [\ln(Q^2 / 1 $GeV$^2 ) + 3]$ 
for $Q^2 > 0.05 $GeV$^2$ and $1.08$ below this. 
This gluon density is then fitted to the available data on
$F_2$. Predictions for the diffractive cross section (which is
proportional to $ [\psi (x_{\Pomm},k_T^2,Q_0^2)]^2 $ integrated over $k_T^2$)
with $q \bar q$ and $q \bar q g$ in final state are then 
presented over a wide range of $\beta$. Now the relevant scale in 
$\alpha_{\Pomm} $ is the virtuality $k_T^2$.

In the limit $\beta \rightarrow 1$ the longitudinal contribution,
which is formally `higher twist', is
finite so is expected to dominate over the transverse
part which goes like 1 - $\beta$. 
This highlights the fact that the concept 
of `twist' must be applied very carefully in diffraction 
- contributions which naively appear higher twist may in
fact dominate at high $Q^2$ in certain regimes.
With an additional gluon in the
final state one finds a $(1-\beta)^3$ behaviour at large $\beta$. For small
$\beta$ this configuration dominates and the cross section diverges
like $1/\beta $. In summary, a characteristic $\beta$ spectrum is found that
shows that emission of the additional gluon is bound to the small $\beta$
region whereas the large $\beta$ is dominated by the longitudinal photon.
Numerical results, labelled `BW', using the ansatz of eq.(\ref{eq:bw}) for $F_2^D$,
and $R^D$ as function of $\beta$ and $Q^2$ are shown in 
figs.(\ref{fig:F2},\ref{fig:R}).

The large mass, small $\beta$ or triple Regge regime ($s \gg M_x^2 \gg
Q^2 \gg \Lambda^2_{QCD} $) has also been investigated in detail by Bartels 
and  W\"{u}sthoff (see \cite{BW2,W,LW} and refs). Theoretically the
emergence of a 4 gluon t-channel state which builds up the large
diffractive mass is expected. 
Experimentally, this region is hard to investigate since 
the requirement of a large mass tends to
close up the rapidity gap making it difficult to distinguish
experimentally from the non-diffractive background and also because the
diffractive final state may not be fully contained in the main detector.
This situation is improving now that the first data collected with the LPS 
is becoming available \cite{zeuslps}. 
For the purpose of this report we will 
discuss expectations in the not-too-small $\beta$ regime.

\begin{figure}[htb] 
\vspace{2mm}
\begin{center}
\leavevmode
\hbox{%
\epsfxsize=6in
\epsfysize=6in
\epsffile{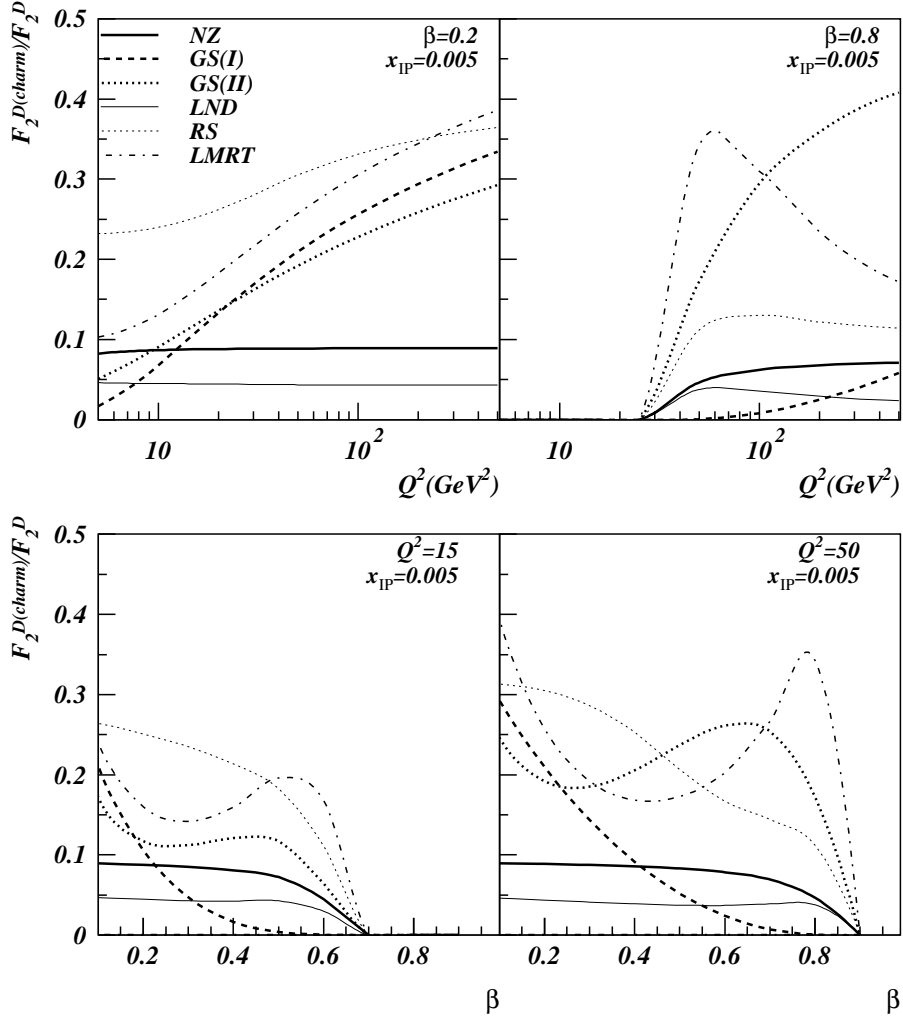}}
\end{center}
 \caption[junk]{{\it Predictions for the charm content in diffraction,
     as a fraction on the total diffractive sample, as a function of
     $\beta$ and $Q^2$ for $x_{\Pomm}= 0.005$. 
     The maximum value of $\beta$  reflects the charm
     threshold and increases with $Q^2$.}}
\label{fig:FC} 
\end{figure}

Diehl \cite{MD} has calculated the contribution of $q \bar q$ in the
final state to the diffraction cross section in the
non-perturbative   two gluon exchange model of Nachtmann and 
Landshoff \cite{LN,DL2}. Numerical predictions for this model 
(applicable for not-too-small $\beta$), labelled `LND',  are shown in 
figs.(\ref{fig:F2},\ref{fig:R},\ref{fig:FC}). This model predicts a
relatively small contribution of charm in diffraction (less than 10 \%
over a wide range of $x_{\Pomm},\beta, Q^2$).

The high $\beta$, small mass regime of diffraction is considered explicitly
in \cite{GLM} who work in co-ordinate space of the transverse distance
between the quark and the anti-quark. 
They claim that at high enough $\beta$ ($\geq 0.4$)
only  the $q {\bar q}$ contributes (in agreement with \cite{BW1}) 
and that for $\beta \geq 0.7$
diffractive scattering from the longitudinal photon dominates 
for which only small distances ($b_T \sim 1/Q $) contribute. 
The effective scale of the gluon density relevant to 
diffraction is $k^2_T/ (1 - \beta)$ 
(see also \cite{BLW}) which is clearly hard for high $\beta$. 
This implies that for high $\beta$ (see fig.(\ref{fig:R})) 
$R^D$ becomes greater than unity in sharp contrast to the Pomeron 
structure function model of \cite{GBK1}.
For the transverse photon distances of $b_T \sim 1 $GeV$^2$ dominate which is
used to justify the use of perturbative QCD and the use of evolution equations,
using GRV input distributions, for the diffractive structure functions.

The series of papers by Genovese, Nikolaev and Zakharov  
\cite{GNZ1,GNZ2,GNZ3,NZ3} provides a model for
diffraction inspired by the QCD dipole approach \cite{MUELLER1,NZ1,NZ2} 
to the generalised BFKL \cite{BFKL} equation. 
In \cite{GNZ3} they strongly reject the factorizable pomeron
model and instead suggest that a two component structure function for the
pomeron with valence and sea partons having different pomeron flux
factors. The absolute normalizations of these components of the
diffractive structure function are substantially the same as
evaluated in 1991 \cite{NZ1}, before the HERA data have become available.
In recent papers for this Regge factorization breaking model
specific predictions for $F_L$ \cite{GNZ1}  and charm \cite{GNZ2} 
are given (see `NZ' in figs.(\ref{fig:F2},\ref{fig:R},\ref{fig:FC})).

The curves, labelled `RS', shown in
figs.(\ref{fig:F2},\ref{fig:R},\ref{fig:FC}) are from a Monte Carlo 
simulation developed by A.Solano and M.Ryskin,
for the dissociation of the virtual photon to two and three jets \cite{RS}. 
The formulae used are the same as those in the LMRT \cite{LMRT} model but use a
GRV \cite{GRV} gluon distribution and a simplified version of the NLO corrections.

Bialas and Peschanski also  present predictions for hard 
diffraction \cite{PB1,PB2}  based on the QCD dipole picture of the BFKL equation. 
In this picture they find that most of the
diffractive cross section comes from the interaction of $q \bar q$
pairs whose transverse size is of the order of the target size as seen
by the virtual photon. The perturbative QCD prediction is enhanced 
by the BFKL resummation and by the number of dipole configurations 
in the initial proton state. In the factorized picture they find
a strong $x_\Pomm$ dependence modified by log corrections. 
They expect $R^D$ to be a strongly varying function of
$\beta$ and to go above unity for large $\beta$. The number of
diffractive events increases with $Q^2$ over the whole range. At small
$\beta$, i.e. large masses, they expect a scaling violation 
to be similar to that seen in $F_2$ at small $x$. Predictions of
this model for $F_2^D$ and $R_D$ 
have been presented recently \cite{ROYON} and are shown, labelled
`BP', in figs.(\ref{fig:F2},\ref{fig:R}).




\section{$F_2^{D~ \mbox{(Charm)}}$}

The ratio of charm events observed in the diffractive structure
function
is in principle a very good test of the hardness of the
processes feeding the $c{\bar c}$ production. 
Clearly a measurement of the $\beta$ and $Q^2$
spectra for these charm events will provide a lot more information. 


If hard QCD dominates in diffraction, i.e. the transverse momenta of
the $q \bar q$ in the
loop are large,  $k_T^2 \sim Q^2$,  the relative yield
of charm in diffraction is
determined by the electric charges of the quarks and should be about
40 \%. In the Pomeron structure function models of \cite{GS} the charm
contribution comes from boson gluon fusion and is indeed large. Model
2 predicts that it should also be large at high $\beta$ in comparison
to model 1 (compare `GS(I)' and `GS(II)' in fig.(\ref{fig:FC})). 
Also since diffraction is a higher twist effect one would
expect the $F^{D(\mbox{charm})} / F^{\mbox{(total charm)}}$ 
to decrease quickly as a function of $Q^2$.


In the naive AJM, since the quark transverse momenta are small, one would expect
a very small charm content. Within
the QCD-improved AJM this may be expected to increase with $Q^2$ and for
sufficiently high $Q^2$ the charm contribution to diffraction 
should approach that anticipated from hard physics.  

The early paper of Nikolaev and Zakharov \cite{NZ3}, predicts that the
diffractive contribution to open charm is around 1 \%.
In a recent paper \cite{GNZ2}, they present predictions for the charm 
contribution to diffraction and suggest a very steep rise at small
$x_{\Pomm}$ strongly breaking Regge factorization; at $x_{\Pomm} =
0.005$ 
this leads to a charm content of about 10 \% 
(see `NZ' in fig.(\ref{fig:FC})).

In a numerical study of the influence of the small $k_T$ region in the
BFKL equation, in \cite{BLV}, it is shown that the dominant
contribution to diffraction comes from the region of small transverse
quark momenta, even for large $Q^2$. 
This would seem to favour a small charm contribution in this model.

The LMRT approach \cite{LMRT} is based on the same Feynman 
graphs for $\gamma\to q\bar{q}$ and $\gamma\to q\bar{q}g$ 
dissociation as \cite{BW1,GLM,BW2,W}
and \cite{GLM,NZ1,GNZ1,GNZ2,GNZ3,NZ3,NZ2}. 
\footnote{these perturbative QCD formulae were first derived in
\cite{NZ1} for $\gamma\to q\bar{q}$ and in \cite{LW} for $\gamma\to q\bar{q}g$}
However in the LMRT case the most realistic MRS(A')  gluon
distribution (which fits all the present data) were used and the main
NLO corrections, including an estimate of the K-factor in the
$O(\frac{\alpha_s}\pi \pi^2)$ approximation, 
were taken into account. Thanks to the large anomalous dimension
$\gamma$ of the gluon structure function $g(x,k_T^2)\propto
(k_T^2)^\gamma$ at small $x=x_{\Pomm}$ 
the infrared divergence is absent from the $k_T$-integral
and, even for the transverse part originated by the light quarks, 
the dominant contribution comes mainly from small distances (see also
\cite{GLM}) and 
doesn't depend too much on the value of the infrared cutoff.
This short distance dominance is reflected in the large charm content
of the Monte Carlo \cite{RS} and of
\cite{LMRT} (see curves 'RS' and 'LMRT' in fig.(\ref{fig:FC}), respectively).
The LMRT predictions are
 normalised using a phenomenological fit to the '93 ZEUS data and show a 
 significant threshold behaviour for $\beta$ approaching the kinematical limit.
 The sharp increase for low values of $\beta$ comes from the inclusion of real 
 gluon emission (see \cite{LMRT}), which is not taken into account in LND.

The measurement of the charm contribution in diffraction, 
which should be available in the near
future (at least for $D^*$ production \cite{CHARM}),
will certainly help our understanding of the interplay of hard
and soft physics in diffraction.



\section*{Acknowledgments}

We would like to thank H. Abramowicz, 
J. Bartels, W. Buchm\"{u}ller, L. Frankfurt, H. Jung and M. Ryskin for
discussions and suggestions for this report. We're also grateful to 
M. Diehl, T. Gehrmann, K. Golec-Biernat, N. Nikolaev, 
C. Royon, A. Solano, M. W\"{u}sthoff for providing numbers for the 
figures at short notice.


\end{document}